**Title:**

# Tailoring hBN's Phonon Polaritons with the Plasmonic Phase-Change Material In$_3$SbTe$_2$

Author(s) and Corresponding Author(s)*:

Lina Jäckering*,+,[1], Aaron Moos[+,1], Lukas Conrads[1], Yiheng Li[1], Alexander Rothstein[2,3], Dominique Malik[1], Kenji Watanabe[4], Takashi Taniguchi[5], Matthias Wuttig[1,6,7], Christoph Stampfer[2,3,7], Thomas Taubner*,[1]

*jaeckering@physik.rwth-aachen.de; taubner@physik.rwth-aachen.de

[+]both authors contributed equally

**Affiliations**

[1]1st Institute of Physics (IA), RWTH Aachen University, 52074 Aachen, Germany

[2]2nd Institute of Physics, RWTH Aachen University, 52074 Aachen, Germany

[3]Peter Grünberg Institut (PGI-9), Forschungszentrum Jülich, 52425 Jülich, Germany

[4]Research Center for Electronic and Optical Materials, National Institute for Materials Science, 1-1 Namiki, Tsukuba 305-0044, Japan

[5]International Center for Materials Nanoarchitectonics, National Institute for Materials Science, 1-1 Namiki, Tsukuba 305-0044, Japan

[6] Peter Grünberg Institut (PGI-10), Forschungszentrum Jülich, 52425 Jülich, Germany

[7]Juelich-Aachen Research Alliance (JARA-FIT), 52425 Juelich, Germany

**Abstract:**

Polaritons in van-der-Waals materials (vdWM) promise high confinement and multiple tailoring options by optical structures, e.g., resonators, launching structures and lenses. These optical structures are conventionally fabricated using cumbersome multi-process lithography techniques. In contrast, phase-change materials (PCMs) offer fast and reconfigurable programming of optical structures. PCMs can reversibly be switched between two stable phases with distinct permittivities by local heating, e.g., by optical laser pulses. While the well-known dielectric PCM GeSbTe-alloys feature only a permittivity change, the PCM $In_3SbTe_2$ can be switched between a dielectric and metallic phase. This makes $In_3SbTe_2$ promising for programming metallic launching structures. Here, we demonstrate direct optical programming and thereby rapid prototyping of optical launching structures in $In_3SbTe_2$ to tailor and confine polaritons in vdWM. We combine the vdWM hexagonal boron nitride (hBN) with $In_3SbTe_2$ and optically program circular resonators for hBN's phonon polaritons through hBN into $In_3SbTe_2$. We investigate the polariton resonators with near-field optical microscopy. Demonstrating the reconfigurability, we decrease the resonator diameter to increase the polariton confinement. Finally, we fabricate focusing structures for hBN's phonon polaritons whose focal point is changed in a second post-processing step. We promote $In_3SbTe_2$ as a versatile platform for rapid prototyping of polariton optics in vdWM.

**Introduction**

Many nanophotonic devices are based on the hybridization of photons with polarization charges in a material, e.g. plasmons in a metal or phonons in a polar crystal. This hybridization results in a confined quasiparticle called polariton.[1] In recent years, interest in nanophotonics has been drawn to van der Waals materials (vdWM) as they host polaritons with high confinement due to their two-dimensionality, with long propagation length and they can easily be combined to heterostructures.[2,3]

Improving nanophotonic devices requires advanced manipulation of polariton propagation and their enhanced confinements.[3] The polariton propagation can be manipulated by their coupling to optical structures, e.g., resonators[4,5], or by changing the dielectric environment[6–8]. The achieved manipulation of propagation allows for optical waveguiding and lensing whereas a high confinement by optical resonator structures is interesting for molecular sensing.[2,3] However, most of the optical structures rely on cumbersome fabrication techniques.

Conventionally, micrometer-sized optical structures for polaritons in vdWM are either fabricated by directly etching the vdWM or by precisely aligning the vdWM flake with a pre-defined structure of, e.g., gold. Both fabrication options require several process steps including flake transfer, careful alignment of the flake, e-beam or focused ion beam, lithography, and etching.[4,5,9–11] Among those techniques are several that require the access to clean room facilities. Apart from being time-consuming, a further drawback of conventional fabrication techniques is that -once fabricated- the structures cannot be changed or adapted.[4,5,9–11] Relying on conventional fabrication techniques hinders fast prototyping and adaption of optical structures for polaritons.

Fast prototyping of reconfigurable micrometer-sized structures can be realized with phase-change materials (PCMs). PCMs offer non-volatile switching between two (meta-) stable phases, the amorphous and the crystalline phase, which exhibit distinct permittivities.[12,13] The reversible phase change can be induced by local heating of the PCM, e.g., by electrical or optical pulses.[12] A pulsed laser can be used to locally switch the PCM and thereby optically program mircometer-sized structures. In contrast to the conventional fabrication techniques, only a fast two-step process is required, i.e. the vdWM flake has to be exfoliated onto the PCM and the optical structure has to be programmed, no clean-room techniques are needed, and the structures are reconfigurable.

In the past, most applications of PCMs in the field of nanophotonics relied on a combination of a bulk, surface polariton hosting material with the well-known PCM $Ge_3Sb_2Te_6$ (GST). GST has two dielectric phases and therefore allows for a change of the dielectric environment.[12] This change of the dielectric environment has been exploited to investigate switchable and reconfigurable surface phonon polariton (SPhP) resonators.[14,15] The PCM GST has also been introduced to polaritonics with vdWM by exploiting the permittivity change of GST to modulate the propagation of hyperbolic phonon polaritons (HPhPs) in hexagonal boron nitride (hBN). For example, planar lenses have been programmed by locally crystallizing GST and thereby providing a permittivity change between amorphous and crystalline GST of about 25 from 12 to 37 for their real parts between 1350 and 1650 cm$^{-1}$.[6]

In contrast, the plasmonic PCM $In_3SbTe_2$ (IST) offers a phase change between the dielectric amorphous ($\varepsilon > 0$) and the metallic crystalline phase ($\varepsilon < 0$) making it a promising platform for reconfigurable polariton launching structures.[16–18] Previously, IST has been extensively studied for antenna resonance tuning, tailoring thermal emission and even for programming large-area infrared beam-shaping metasurfaces.[19–25] Recently, its capability to confine SPhPs into cavities directly written into IST has been demonstrated for the SPhPs hosting bulk material SiC.[17] In contrast to the surface confined SPhPs in SiC, the HPhPs in hBN propagate through the hBN-slab forming volume-confined modes. This raises the question whether the HPhPs can be modified sufficiently by the two phases of IST. Recently, the efficient launching of HPhPs in hBN at a phase boundary of IST has been demonstrated - owing to the huge permittivity change from 14 to -133 of 147 in their real part between 1350 and 1650 cm$^{-1}$ when switching IST from the dielectric amorphous to the metallic crystalline phase.[8] As the crystalline phase of IST is metallic it provides a high field confinement.[8] However, resonators imprinted in PCMs for polaritons in vdWM have not been investigated. Here, we continue exploring the arising platform for the versatile tuning of polaritons in vdWM with IST. We study the confinement of hBN's HPhPs into a circular resonator directly written into the IST after having placed an hBN flake on it. Further, we demonstrate the capability of reconfiguring the resonator size which allows, for example, to increase the field confinement. Finally, we show polariton focusing with an optically programmed arc structure. In a subsequent post-processing adaptation of the arc structure we shift the focus point.

**Results**

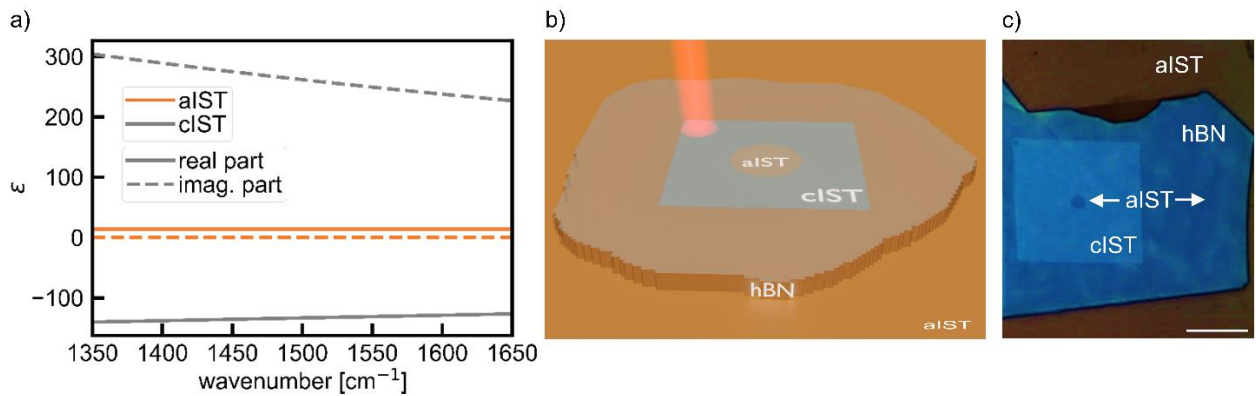

*Figure 1: Direct programming of hBN phonon polariton resonators into IST.* a) Real (solid) and imaginary (dashed) part of the permittivity of IST's dielectric amorphous (orange) and metallic crystalline (grey) phase. b) Sketch of the laser crystallization process of the PCM IST with which an amorphous resonator structure is programmed into the IST below hBN. c) Light microscope image of the rectangular laser-crystallized field in the IST below the 97 nm thick hBN flake (blue) leaving a circular amorphous resonator in the center. The scale bar is 10 μm.

We design circular resonators by exploiting the two reversibly switchable phases of IST. While the amorphous phase has a slightly positive real part of the permittivity, the crystalline phase has a strongly negative real part following a Drude-like behavior and thereby demonstrating its metallic behavior (c.f. Figure 1a). We crystallize the amorphous IST below the hBN-flake ($d_{hBN}$=97 nm) with precise laser pulses to create a circular amorphous resonator (c.f. sketch in Figure 1b and light microscope image in Figure 1c). To crystallize the IST we use a pulsed laser diode with a wavelength of 660 nm. At this wavelength the hBN is highly transparent. This allows us to create laser-crystallized spots through the hBN in the IST below. Due to the elliptical beam profile, we create elliptically crystallized spots that we spatially overlay to program an amorphous resonator. The time to write the field of crystalline laser spots leaving an amorphous resonator (see Figure 1c) takes about 10 minutes. This fast prototyping allows us to study resonators for hBN flakes of various thicknesses. We find that we can optically program resonators into the IST through hBN flakes with thicknesses ranging from 35 to 465 nm (see Supplementary Note 1).

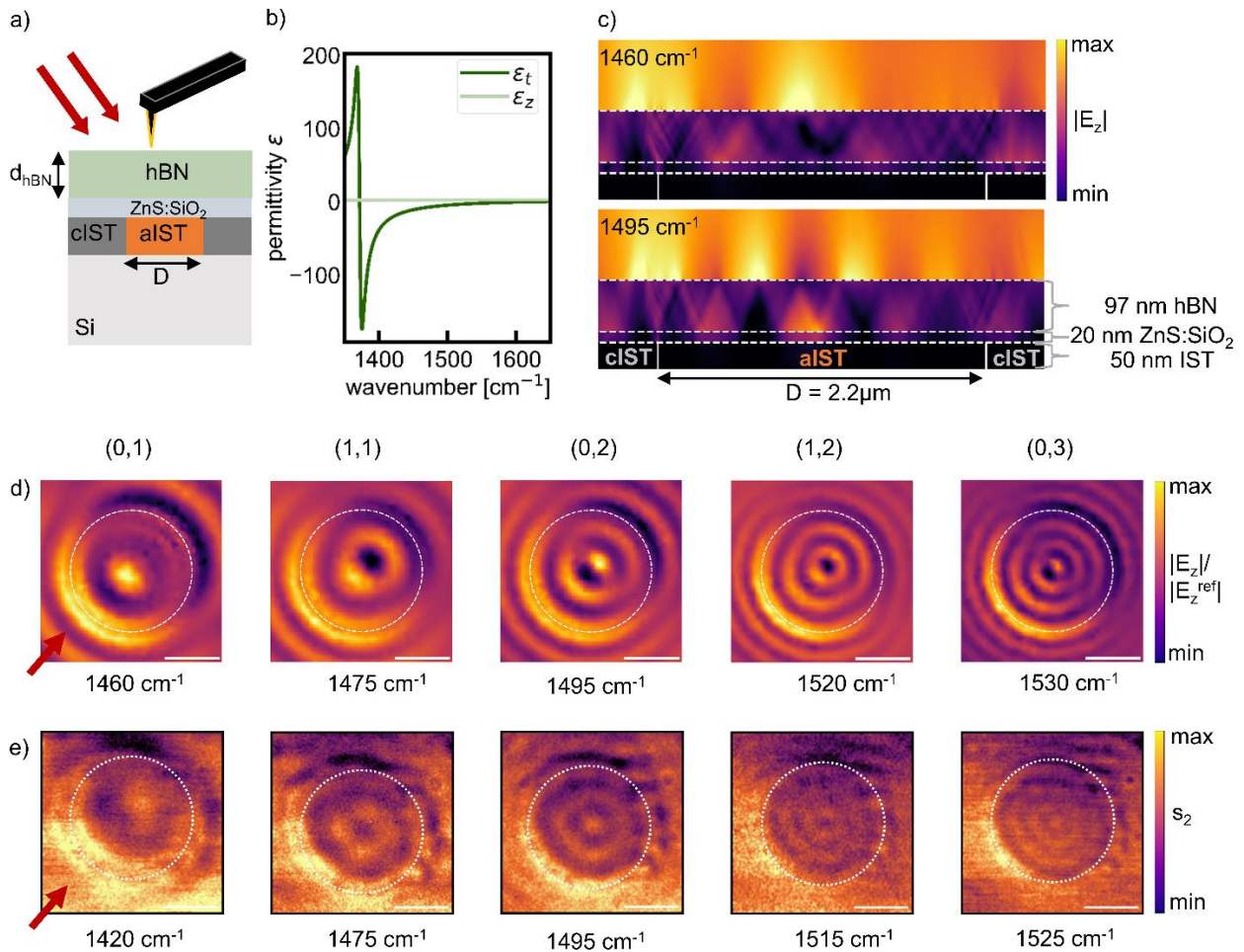

*Figure 2: Direct imaging of the resonator modes of hBN phonon polariton resonators programmed into IST.* a) Sketch of the s-SNOM imaging of modes in the amorphous resonator structure for HPhPs in hBN ($d_{hBN}$ = 97 nm) written into IST as a side view. b) In-plane (dark

*green) and out-of-plane (light green) permittivity of hBN in the investigated upper reststrahlenband taken from ref.[26]. c) Cross section of the simulated HPhP mode behavior in the hBN above the amorphous resonator structure (D = 2.2 µm). d) Simulations of the electric field of the resonator modes at five different frequencies showing various resonator modes. e) s-SNOM amplitude images of hBN's HPhPs at the same modes as in d). The red arrows in d) and e) indicate the in-plane illumination direction of the incident laser. The scale bars correspond to 1 µm.*

First, we investigate the resonator modes of hBN's HPhPs in the upper reststrahlenband with scattering-type scanning near-field microscopy (s-SNOM) which is sensitive to local near-fields, e.g. propagating and localized polaritons, and enables a wavelength-independent resolution down to 10 nm.[27] In s-SNOM, a sharp tip is brought into proximity to the sample surface and illuminated with laser light (c.f. Figure 2a). Acting as an optical antenna, the tip strongly enhances the electric fields. The scattered field is detected interferometrically to obtain the amplitude $s_n$ and the phase $\varphi_n$, here n indicates the demodulation order.

The polaritonic near-fields at the hBN surface that we can image with s-SNOM (see sketch in Figure 2a) originate from hBN's volume-confined phonon polaritons and are tailored by the IST below. hBN has an anisotropic crystal structure as it has polar bonds in-plane but van-der Waals bonds out-of-plane. Therefore, hBN has different permittivities along the in- and out-of-plane axes (shown in Figure 2b), which are not only different but also opposite in sign. This leads to hyperbolic isofrequency curves for hBN's hyperbolic phonon polaritons (see Supplementary Note 2) and defines hBN as a natural hyperbolic material.[28–30] The HPhPs show a propagation behavior distinct from polaritons in isotropic materials. While in isotropic materials the photon wavenumber defines the magnitude of the polariton wavevector, in hyperbolic materials the magnitude of the polariton wavevector can be arbitrarily high but the angle of polariton propagation is defined. The HPhPs propagate under a restricted angle through the hBN slab and are reflected at the interfaces of the slab. The HPhPs form volume-confined modes similar to waveguide modes[28,30] (see Supplementary Note 2 for a detailed introduction to hBN's HPhPs). Further, the HPhPs in hBN feature the hyperlensing effect allowing for super-resolution and enlarged imaging of structures below the hBN.[10,30–32]

As introduced in Figure 1, we now combine hBN with IST that can be switched between two phases with distinct permittivities. By changing the permittivity of the material below hBN, we can modify the propagation behavior of hBN's HPhP (see Supplementary Note 3). We optically program local amorphous circular structures into IST which serve as resonators for the HPhPs propagating in the hBN above the IST. In the used layer stack (see Figure 2a) the IST is capped by a layer of ZnS:SiO$_2$ to prevent the IST from oxidation.

The electric fields simulated (see Methods) in a cross section across the hBN IST heterostructure in Figure 2c demonstrate that the HPhPs are confined within the hBN slab and propagate only under a restricted angle. The simulations show that the polaritons are launched at the boundary between crystalline and amorphous IST and propagate in both directions away from the boundary. For increasing frequency (bottom in Figure 2c) the propagation angle of the confined polaritons decreases.[10,30] The confined polariton propagation within the hBN slab transfers into a periodic pattern of high fields above the hBN slab. In a top view, this pattern of high fields corresponds to the appearance of rings of high fields due to the circularly symmetric geometry of the resonator (c.f. simulations in Figure 2d). The field patterns in our simulations are asymmetric due to the considered illumination direction from the left bottom marked by the red arrow that resembles the illumination direction of the s-SNOM measurements shown in Figure 2e. At 1460 cm$^{-1}$, we observe a single spot of high fields in the center of the resonator. With increasing frequency at 1475 cm$^{-1}$, this spot evolves into a ring of high fields with a minimum in its center. For further increasing frequencies (1495 -1530 cm$^{-1}$), we observe alternating minima and maxima in the center of the resonator and an increasing number of rings of high fields. The observation of rings of high fields with alternating minima and maxima in the center is in agreement with results on circular resonators for SPhPs and interpreted as the evolution of modes in this resonator.[14,17,33] Theoretically, the modes of such circular resonators for surface polaritons can be calculated by $k_p D + \phi = 2x_n(J_m)$, where k$_p$ is the polariton wavevector, D the resonator diameter, ϕ the reflection phase at the resonator boundaries and $x_n(J_m)$ the n-th root of the Bessel function of first kind and order m $J_m$.[34] Consequently, the observed modes are denoted as (n,m). We attribute the observed evolution of modes in the resonator for the HPhPs to the evolution from the (0,1) mode at 1460 cm$^{-1}$ to the (0,3) mode at 1530 cm$^{-1}$.

Our s-SNOM amplitude images (Figure 2e) validate this confinement of the polaritons experimentally as the s-SNOM amplitude images of the resonator at different frequencies between 1420 and 1525 cm$^{-1}$ reveal the evolution of the same resonator modes from (0,1) to (0,3). We find a good agreement with the simulated field patterns. Thereby, we demonstrate direct programming of resonator structures for hBN's HPhPs into IST, extending the recently opened platform of programming resonators into IST for surface PhPs in bulk materials[17] to volume-confined hyperbolic PhPs in vdWM.

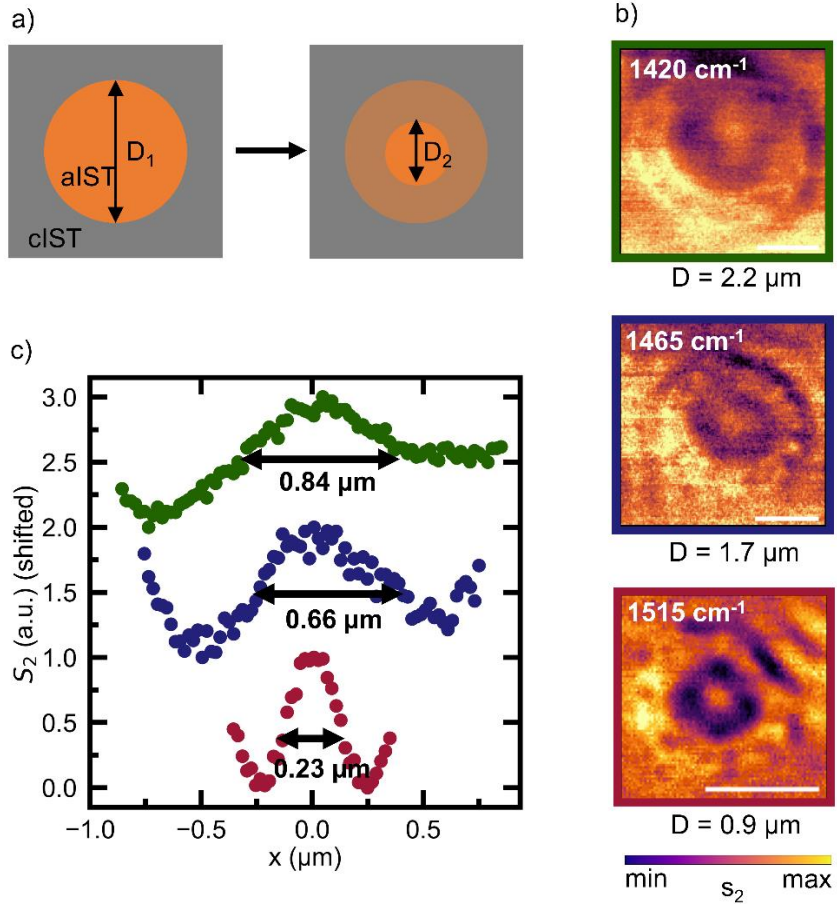

*Figure 3: Reconfiguring the resonator diameter allows for tailoring the polariton confinement:* a) By laser crystallization the diameter ($D_1$) of the programmed resonator (left) is decreased ($D_2$). b) s-SNOM amplitude images of the (0,1) resonator mode in an amorphous resonator whose diameter is decreased sequentially from D = 2.2 µm (top, green) to D = 0.9 µm (bottom, red). The scale bars correspond to 1 µm. c) Line profiles of the resonator modes in b) reveal the increasing confinement of the mode for a decreasing resonator diameter. The line profiles are shifted for better visibility. The arrows and values correspond to the full-width-half-maximum.

Next, we reconfigure the resonator diameter by additional laser crystallization as sketched in Figure 3a. We sequentially decrease the diameter from D = 2.2 µm to D = 1.7 µm and finally D = 0.9 µm. With s-SNOM we visualize the (0,1) mode where we observe a single maximum in the center of the resonator for each diameter that appears at higher wavenumbers for decreasing diameters (c.f. Figure 3b). By decreasing the diameter, the full-width-half-maximum (FWHM) of the peak in the center decreases from 0.84 µm to 0.23 µm (c.f. Figure 3c), demonstrating a strong confinement of the polaritons in the amorphous resonator structure written into IST. We find an enhancement of the confinement $\frac{\lambda_0}{FWHM}$ from 8 for the largest resonator (D = 2.2 µm) to 31 for the smallest resonator (D = 0.9 µm).

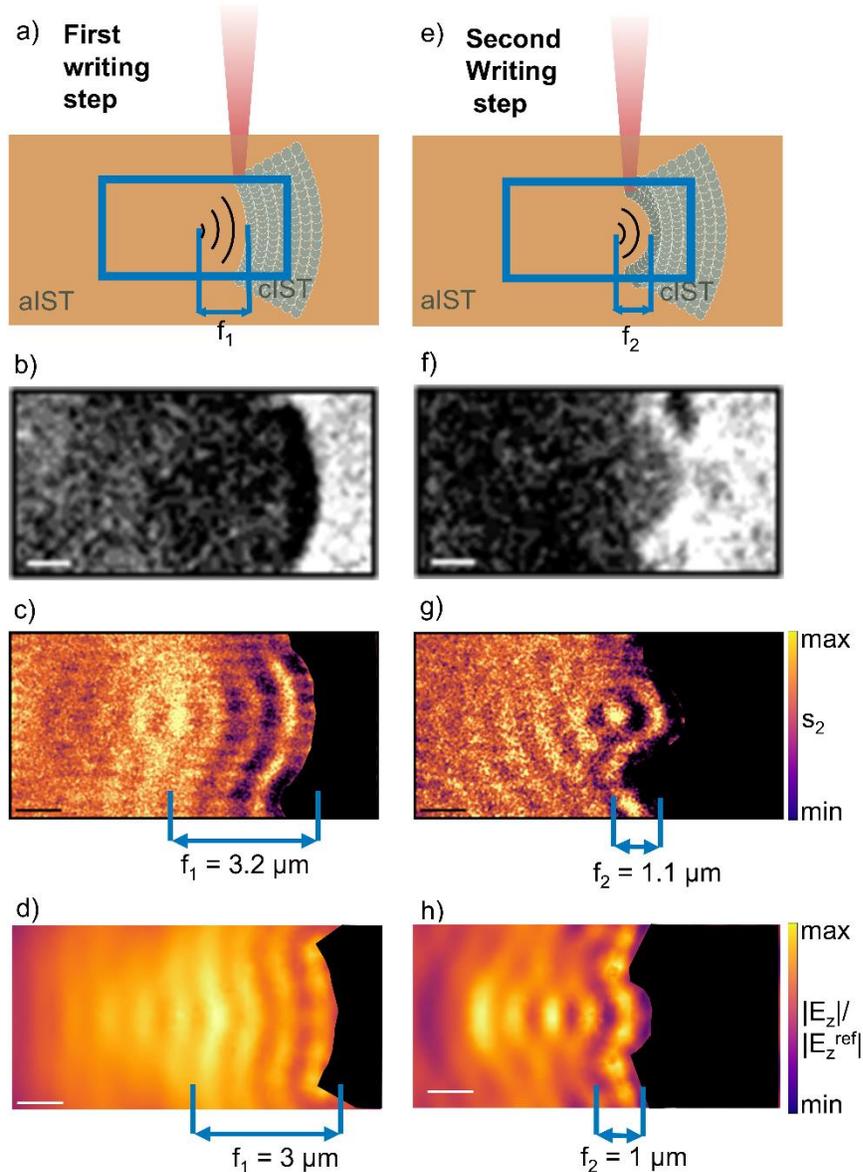

*Figure 4: Reconfigurable focusing of phonon polaritons in hBN. a) Sketch of a launching arc structure directly written into IST by laser crystallization. The black lines indicate the wavefronts of HPhPs launched by the structure. b) Light microscope image of the crystalline launching arc structure below hBN corresponding to the area marked by the blue rectangle in a). s-SNOM amplitude image taken at 1480 cm$^{-1}$ c) and electric field simulation d) of hBN's HPhPs launched and focused to 3 µm by the launching arc structure. e) Sketch of reconfiguring the arc structure by additional laser crystallization of the IST. The black lines indicate the wavefronts of HPhPs launched by the structure. f) Light microscope image of the reconfigured arc structure with a decreased curvature radius corresponding to the area marked by the blue rectangle in e).*

*s-SNOM amplitude image taken at 1480 cm$^{-1}$ g) and electric field simulation h) of hBN's HPhPs launched and focused to 1 µm by the reconfigured launching arc structure. The scale bars correspond to 1 µm.*

Finally, we program a crystalline arc structure into the IST below a 102 nm thick hBN flake as sketched in Figure 4a. The arc structure is designed to focus hBN's HPhPs as the black sketched wave fronts indicate. Our design was inspired by Huber et al.[35] who fabricated a gold arc structure on SiC by conventional fabrication techniques to focus the SPhPs in SiC. The optical light microscope image (Figure 4b) of the area marked with a blue rectangle in Figure 4a shows a bright arc structure corresponding to the crystallized area. This arc structure is designed with an inner radius of 3 µm such that it focusses the HPhPs onto a focal point with a focal length of 3 µm. The s-SNOM amplitude image in Figure 4c reveals a fringe pattern that tapers towards the left until a focal point at about 3.2 µm distance from the arc structure is observed. Further to the left the fringe pattern widens and flattens out. The fringe pattern arises from the interference of the incident light and the HPhPs in hBN launched at the boundary of the crystalline IST structure. The electric field simulation in Figure 4d supports our experimental observations.

Now, we aim for a change of the focal length towards 1 µm. Therefore, we perform a 2$^{nd}$, post-processing writing process as sketched in Figure 4e. We now change the inner curvature radius of the arc structure to 1 µm such that the HPhPs are focused to a point closer to the launching structure as indicated by the sketched wavefronts. The optical light microscope image in Figure 4f shows the modified crystalline arc structure with a smaller curvature radius. The s-SNOM amplitude image in Figure 4g reveals a bright fringe close to the arc structure and a bright spot at a distance of about 1.1 µm corresponding to hBN's HPhPs that are now focused to a focal point in 1.1 µm distance. The experimentally observed fringe pattern agrees well with the electrical field simulation (c.f. Figure 4h). By reconfiguring the crystalline arc structure, we have adapted the focal length of our focusing structure within a single post-processing step. The easy reconfigurability makes IST a versatile and flexible platform to study polariton modification by optical launching structures.

**Conclusion**

We promote the phase-change material $In_3SbTe_2$ as a versatile platform for studying the intriguing properties of polaritons in vdWM such anisotropic propagation and high confinement. We demonstrate circular resonator structures directly written into IST below hBN by laser crystallization. The confined resonator modes are visualized with near-field optical microscopy. We decrease the resonator diameter and thereby increase the mode confinement to $\frac{\lambda_0}{31}$ to showcase the reconfigurability of the programmed structures. The direct programming of resonator structures enables fast prototyping that we exploit to program resonators for hBN thicknesses ranging from 35 to 465 nm (see SI). For all thicknesses the resonator modes can be identified demonstrating the versatility of the hBN IST platform. Finally, we program a crystalline focussing structure with a focal length of 3.2 µm into the IST in order to focus the HPhPs. By post-processing, meaning a further crystallization process, we adapt the focal length to 1.1 µm.

Optical structures written into IST are not limited to resonators. In principle, much more complex launching structures such as metasurfaces to steer polaritons can be programmed into IST similar to those written into GST[6]. As IST allows for fast prototyping of launching structures through vdWM of various thicknesses, we expect IST to be a versatile substrate where optical structures for polaritons in vdWM can be directly imprinted. With the method demonstrated we shrink down the time-consuming multi-step fabrication process of optical structures to a fast two-step process, i.e. the exfoliation of the vdWM onto the IST and the subsequent optical programming of the desired structure into the IST. Thereby, IST can facilitate the exploration of polaritons in other exciting in-plane anisotropic vdWM such as α-$MoO_3$[36,37], β-$Ga_2O_3$[38] and their heterostructures. Apart from reconfiguring the programmed structures, a second modulation of the polaritons could be added by stacking the phonon polariton hosting materials into heterostructures with graphene since the resulting hybrid phonon plasmon polaritons are gate tunable[39].

**Methods**

**Sample Fabrication.** Using direct current and radio frequency sputter deposition, a 50nm thick layer of amorphous $In_3SbTe_2$ and a 20nm thick capping layer of $ZnS:SiO_2$ were applied on a Si substrate. The capping layer (80% ZnS, 20% $SiO_2$) prevents the sample from oxidation. Flakes of hBN were then directly exfoliated onto the previously described amorphous PCM sample.

**Optical switching.** A home-built laser switching setup was used to crystallize the PCM locally. Here, the light of a 660 nm laser diode is focused on the sample with a 10-fold objective with NA = 0.25. Each individual crystallized spot for the resonator structures was created by applying 100 pulses with a power of 91 mW and a pulse length of 800 ns. Due to the elliptically beam profiles the crystallized spots are elliptical. One elliptical crystallized spot has a lateral extension of approximately 1 x 1.5 µm². The sample was placed on a Thorlabs piezo stage allowing for precise movements of the sample. The amorphous resonator is created by placing multiple crystalline spots next to each other and leaving out a circular amorphous region in the center, the resonator. As the hBN flake we used to investigate the launching arc structures is thicker, we slightly adapted the parameters for laser crystallization. In case of the arc structures, the individual crystallized spots were created by applying 100 pulses with a power of 103 mW and a pulse length of 900 ns. Again, the structures were created by placing multiple spots next to each other.

**s-SNOM.** We used a commercially available s-SNOM (NeaSNOM neaspec GmbH) to image the polaritons in hBN on IST. We combined commercially available quantum cascade lasers (MIRcat-QT Mid-IR Laser and QCL by DRS Daylight Solutions) with a liquid nitrogen-cooled

Mercury Cadmium Tellurium detector to investigate polaritons in the frequency range from 1390 to 1600 cm$^{-1}$. We used the pseudoheterodyne detection scheme to obtain s-SNOM amplitude and phase. We performed our measurements using tapping amplitudes between 65 and 75 nm and demodulated the signal at the second demodulation order.

**Numerical Field Simulations.** Numerical simulations were performed with the commercial solver CST Studio Suite from Dassault Systèmes. The simulation was modelled with Floquet port excitation using an incident angle of the p-polarized light of 60° normal to the surface and unit cell boundary conditions in lateral and open boundary conditions in vertical dimensions.

The constant loss-free refractive index of 3.4 was used for Si. The dielectric function of IST is shown Figure 1a. A constant refractive index of 2.1 was assumed for the ZnS:SiO$_2$. The anisotropic dielectric function of hBN was modelled according to Figure 2b.

The electric field was extracted 1 nm above the sample surface. The fields were normalized with respect to the incident field. Cross-section fields were calculated normal to the p-polarized light in the center of the resonator structure.

Effects of the SNOM tip were not considered in our simulations.

**Code Availability**

The computer codes developed for this study are available from the corresponding author upon request.

**Supporting Information**

Additional materials (PDF)

**Acknowledgements**

The authors thank Maike Kreutz for the sputter deposition of the thin film layer stack. L.J. acknowledges the support of RWTH University through the RWTH Graduate Support scholarship. L.J., L.C. and T.T. acknowledge support by the Deutsche Forschungsgemeinschaft (DFG No. 518913417 & SFB 917 "Nanoswitches"). M.W. acknowledges support by the Deutsche Forschungsgemeinschaft (SFB 917 "Nanoswitches"). A.R. and C.S. acknowledge support from the European Union's Horizon 2020 research and innovation programme under grant agreement no. 881603 (Graphene Flagship), the Deutsche Forschungsgemeinschaft (DFG, German Research Foundation) under Germany's Excellence Strategy - Cluster of Excellence Matter and Light for Quantum Computing (ML4Q) EXC 2004/1-390534769, the FLAG-ERA grant PhotoTBG, by the Deutsche Forschungsgemeinschaft (DFG, German Research Foundation) - 471733165. K.W. and Ta.Ta. acknowledge support from the JSPS KAKENHI (Grant Numbers 20H00354 and 23H02052) and World Premier International Research Center Initiative (WPI), MEXT, Japan.

212121

**Author Contributions Statement**

L.J., L.C. and T.T. conceived the project. A.R. and D.M. fabricated parts of the samples. A.M. performed the s-SNOM measurement. L.J. and Y.L. produced preliminary data performing s-SNOM measurements. L.J. supervised the s-SNOM measurements. A.M. performed the numerical field simulations under supervision of L.C. L.J., A.M. and L.C. analyzed the data. M.W. provided the sputtering equipment and phase-change material expertise. C.S. supervised A.R. on selected aspects of the sample fabrication, based on expertise in 2D material exfoliation and stacking. K.W. and Ta.Ta. provided high-quality hBN crystals. All authors contributed to writing the manuscript.

**Competing Interests Statement**

The authors declare no competing financial interest.

**Supplementary Information:**

**Tailoring hBN's Phonon Polaritons with the Plasmonic Phase-Change Material In$_3$SbTe$_2$**

Author(s) and Corresponding Author(s)*:

Lina Jäckering*,+,1, Aaron Moos+,1, Lukas Conrads[1], Yiheng Li[1], Alexander Rothstein[2,3], Dominique Malik[1], Kenji Watanabe[4], Takashi Taniguchi[5], Matthias Wuttig[1,6,7], Christoph Stampfer[2,3,7], Thomas Taubner*,1

*jaeckering@physik.rwth-aachen.de; taubner@physik.rwth-aachen.de

+both authors contributed equally

**Affiliations**

[1]1st Institute of Physics (IA), RWTH Aachen University, 52074 Aachen, Germany

[2]2nd Institute of Physics, RWTH Aachen University, 52074 Aachen, Germany

[3]Peter Grünberg Institut (PGI-9), Forschungszentrum Jülich, 52425 Jülich, Germany

[4]Research Center for Electronic and Optical Materials, National Institute for Materials Science, 1-1 Namiki, Tsukuba 305-0044, Japan

[5]International Center for Materials Nanoarchitectonics, National Institute for Materials Science, 1-1 Namiki, Tsukuba 305-0044, Japan

[6] Peter Grünberg Institut (PGI-10), Forschungszentrum Jülich, 52425 Jülich, Germany

[7]Juelich-Aachen Research Alliance (JARA-FIT), 52425 Juelich, Germany

**This PDF file includes:**

**Supplementary Note 1: Fast prototyping of resonators written into IST below hBN flakes of various thicknesses**

**Supplementary Note 2: Hyperbolic Phonon Polaritons**

**Supplementary Note 3: Dispersion of Hyperbolic Phonon Polaritons in hBN on amorphous and crystalline IST**

**Supplementary Note 1: Fast prototyping of resonators written into IST below hBN flakes of various thicknesses**

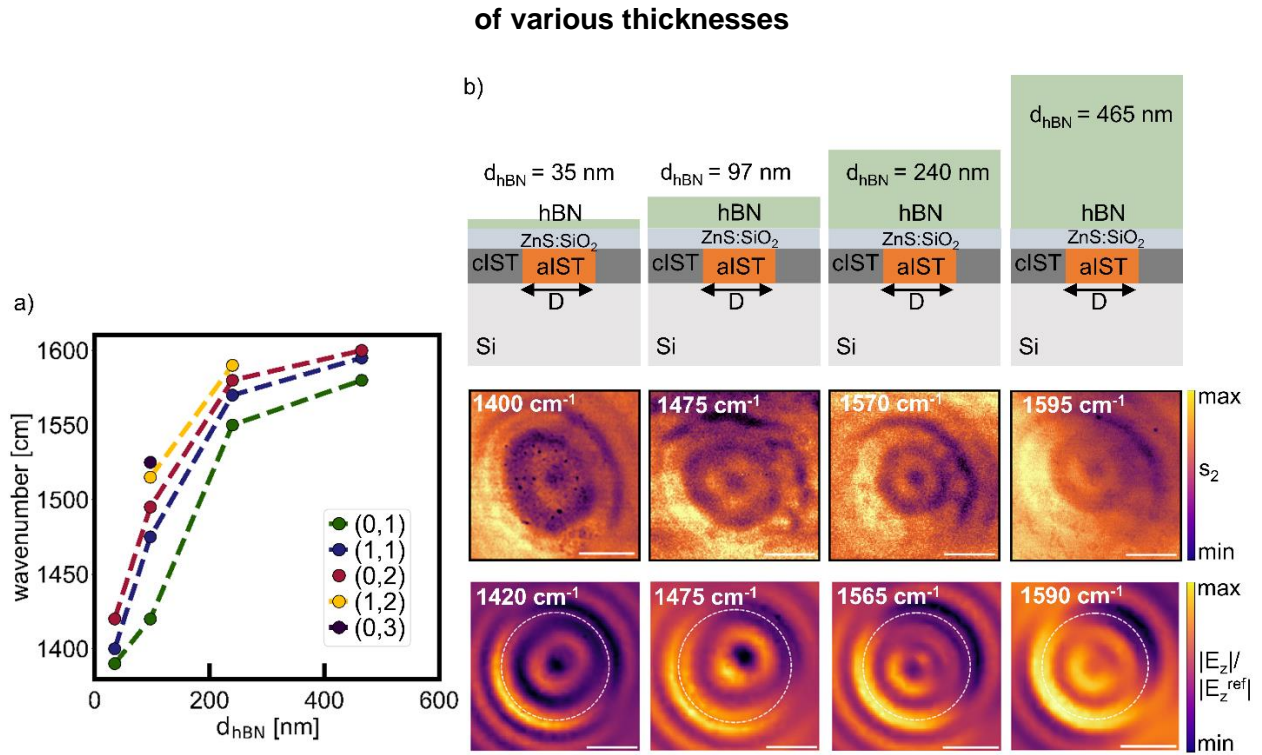

*Figure 2: Fast prototyping of resonators written into IST below hBN flakes of various thicknesses. a) Resonant frequency of various resonator modes for different hBN thicknesses. b) s-SNOM amplitude images (top) and electric field simulations (bottom) of the (1,1) mode (blue in a)) for four different hBN thicknesses. The scale bars correspond to 1 µm.*

Exploiting the fast prototyping we program amorphous resonators (D ~ 2µm) into IST below hBN flakes of four different thicknesses ranging from 35 to 465 nm. Sequential s-SNOM spectroscopy of the resonator modes allows us to identify the mode evolution with illumination frequency for the different hBN thicknesses as summarized in Figure 3a. For each hBN thickness we can at least identify the (0,1), (1,1) and (0,2) mode. With increasing hBN thickness the modes are found at higher frequencies as expected form the thickness-dependent dispersion of hBN's PhP.[1] Figure 3b illustrates the change of the resonance frequency of the (1,1) mode for the different hBN thicknesses showing s-SNOM amplitude images of the mode at the top and corresponding electric field simulations at the bottom.

## Supplementary Note 2: Hyperbolic Phonon Polaritons

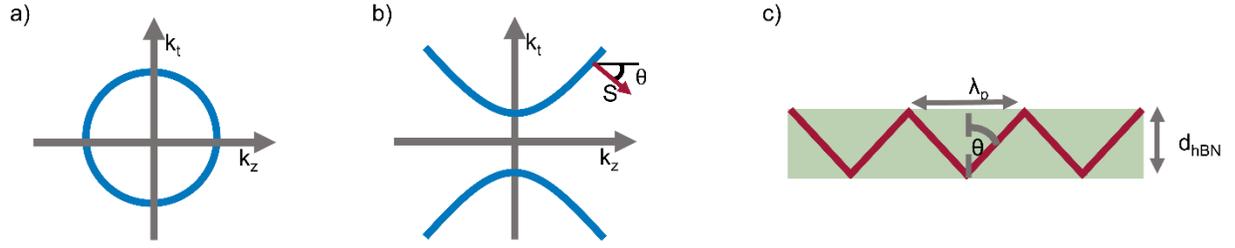

*Figure 2: Hyperbolic phonon polaritons in hBN.* a) Isofrequency curve of an isotropic medium. b) Isofrequency curve of an hyperbolic medium. c) Sketch of the HPhP propagation through an hBN slab.

In isotropic media, the permittivity is the same along all directions and therefore the polaritons propagate uniformly in all directions. The uniform propagation behavior is visualized in the isofrequency curve in Figure 1a. For a given photon wavenumber, the magnitude of the polariton wavevector is fixed whereas the propagation direction is free.

In contrast to isotropic media, hyperbolic media have distinct permittivities along the different optical axes which are opposite in sign.[2] From the dispersion relation, we obtain hyperbolic shaped isofrequency curves as depicted in Figure S2b. For a given photon wavenumber, the magnitude of the polariton wavevector is not fixed as for isotropic media but can -in principle- be infinitely large. The propagation direction, however, is fixed for a given photon wavenumber for large wavevectors. The propagation angle can be approximated by $\tan\theta(\nu) = i\frac{\sqrt{\varepsilon_t(\nu)}}{\sqrt{\varepsilon_z(\nu)}}$.[3,4]

Consequently, inside the hBN-slab the HPhPs propagate under the angle $\theta(\nu)$ as it is sketched in Figure S2c. The HPhPs propagate through the slab and are reflected at the interfaces forming volume-confined modes. The polariton wavelength corresponding to the spacing of the periodic field pattern at the hBN-surface can be derived from the geometry: $\lambda_p = 2\ d_{hBN}\ \tan\theta(\nu)$.[3] When the actual layer stack rather than a free-standing hBN slab is considered, factors such as the reflection phase at the material interface need to be accounted for. Thus, the polariton wavelength will deviate from the simple geometric description.[3]

**Supplementary Note 3: Dispersion of Hyperbolic Phonon Polaritons in hBN on amorphous and crystalline IST**

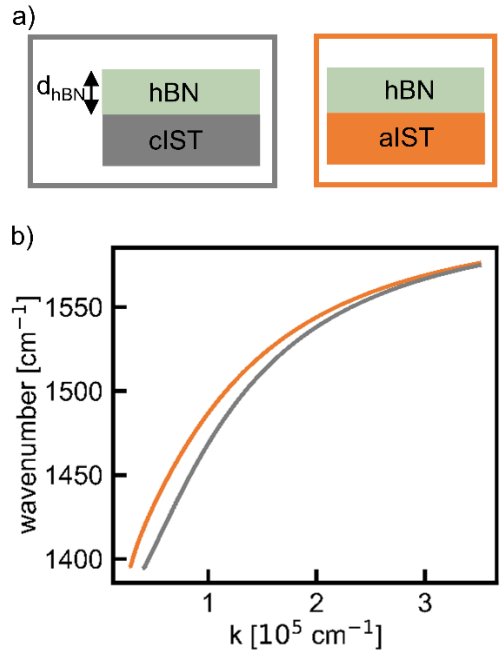

*Figure S3: Dispersion of HPhPs in hBN on aIST and cIST. a) Sketch of the two layer stacks hBN on cIST marked in grey on the left and on aIST marked in orange on the right for which we determined the theoretical dispersion relation in b). b) Calculated dispersion of the HPhPs in hBN on cIST in grey and on aIST in orange.*

The propagation behavoir of the HPhPs in hBN can be altered when switching the crystalline IST below the hBN (sketch on the left of Figure S3a) to amorphous IST (sketch on the right of Figure S3a), as it can be seen in the theoretical dispersion in Figure S3b. We calculated the theoretical dispersion using the Transfer Matrix Method to calculate the reflection coefficient as a function of photon wavenumber and polariton wavevector and then extracted thr mode from the imaginary part of the reflection coefficient. The dispersions of the HPhPs in the two layer stacks show that the HPhPs have a large wavevector (smaller wavelength) when the hBN is on cIST over the whole spectral range. For higher photon wavenumbers the difference between the polariton wavelength decreases and almost vanishes above 1560 cm$^{-1}$.